\documentclass[pre,twocolumn]{revtex4}
\usepackage{amsmath}
\usepackage{amssymb}
\usepackage[dvips]{graphicx}
\usepackage{amsmath,bm,epsfig}
\usepackage{epsfig}
\usepackage{bm}
\newcommand{\B}[1]{{\bm{#1}}}

\newcommand{\pa}{\partial}
\usepackage[latin1]{inputenc}

\begin{document}

\title{Slow rupture of frictional interfaces}

\begin{abstract}
The failure of frictional interfaces and the spatiotemporal structures that accompany it are central to a wide range of geophysical, physical and engineering systems. Recent geophysical and laboratory observations indicated that interfacial failure can be mediated by slow slip rupture phenomena which are distinct from ordinary, earthquake-like, fast rupture. These discoveries have influenced the way we think about frictional motion, yet the nature and properties of slow rupture are not completely understood. We show that slow rupture is an intrinsic and robust property of simple non-monotonic rate-and-state friction laws. It is associated with a new velocity scale $c_{min}$, intrinsically determined by the friction law, below which steady state rupture cannot propagate. We further show that rupture can occur in a continuum of states, spanning a wide range of velocities from $c_{min}$ to elastic wave-speeds, and predict different properties for slow rupture and ordinary fast rupture. Our results are qualitatively consistent with recent high-resolution laboratory experiments and may provide a theoretical framework for understanding slow rupture phenomena along frictional interfaces.
\end{abstract}

\author{Yohai Bar Sinai$^1$, Efim A. Brener$^{1,2}$ and Eran Bouchbinder$^1$}

\affiliation{$^1$Chemical Physics Department, Weizmann Institute of Science, Rehovot 76100, Israel\\
$^2$Peter Gr\"unberg Institut, Forschungszentrum J\"ulich, J\"ulich 52425 Germany}

\maketitle

\section{Introduction}

Understanding the dynamic processes that govern interfacial failure and frictional sliding, e.g. an earthquake along a natural fault, remains a major scientific challenge. Recently, several geophysical and laboratory observations have pointed to the possibility that stress releasing interfacial slip can be mediated by the propagation of rupture fronts whose velocity is much smaller than elastic wave-speeds \citep{Rubinstein2004, Ben-David2010a, Nielsen2010, Peng2010}.

The nature and properties of these slow rupture fronts, and in particular their propagation velocity, are still not fully understood. The experiments of \citet{Rubinstein2004, Ben-David2010a} clearly demonstrate the existence of a minimal propagation velocity below which no fronts are observed. To the best of our knowledge, no theoretical understanding of this minimal velocity is currently available.

Frictional phenomena are commonly described using phenomenological rate-and-state friction models, see for instance \citet{Dieterich1979, Ruina1983, Baumberger2006, Bizzarri2011}.
Two possible mechanisms for generating slow rupture events were invoked in this framework. The first involves a non-monotonic dependence of the steady state frictional resistance on slip velocity \citep{Weeks, Kato2003, Shibazaki2003}, while the second involves spatial variation of frictional parameters and stress heterogeneities \citep{Yoshida2003, Liu2005a}. The former mechanism is an intrinsic property of the friction law, while the latter mechanism is an extrinsic one. The laboratory measurements of \citet{Rubinstein2004, Ben-David2010a, Ben-David2010b}, performed on a quasi-2D spatially homogeneous system, may suggest that the second mechanism is not necessary for the existence of slow rupture.

In this study we show that slow rupture naturally emerges in the framework of spatially homogeneous rate-and-state friction models. Our analysis is based on a friction model that includes an elastic response at small shear stresses and a transition to slip above a threshold stress. The model exhibits a crossover from velocity-weakening friction at small slip rates to velocity-strengthening friction at higher slip rates, which we argue to be a generic feature of friction.

The existence of a minimal rupture front velocity $c_{min}$, which is determined by the friction law and is independent of elastic wave-speeds, is predicted analytically in a quasi-1D limit. We show that there exists a continuum of rupture fronts with velocities ranging from $c_{min}$ to elastic wave-speeds, in qualitative agreement with recent laboratory measurements \citep{Ben-David2010a} and possibly consistent with field observations \citep{Peng2010}. We further show that slow rupture is significantly less spatially localized than ordinary fast rupture. These predictions are corroborated by explicit calculations for a rock (granite) and a polymer (PMMA), demonstrating the existence of slow rupture which is well-separated from ordinary fast rupture. We believe that these results are potentially relevant for slow/silent earthquakes in geological contexts.

\section{A Rate-and-State Friction Model}
\begin{figure}
 \centering
 \noindent\includegraphics[width=1\columnwidth]{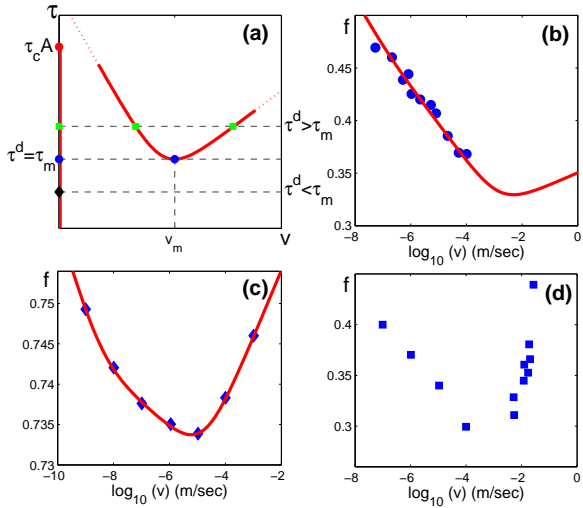}
 \caption{{\bf (a)} A schematic sketch of the homogeneous solutions of $\tau(v)$. {\bf (b)} $f(v)$ for PMMA \citep{Baumberger1999}. The solid line is a fit to Eq. (\ref{steady_sliding}). {\bf (c)} $f(v)$ for granite with $\sigma\!=\!5$ MPa \citep{Kilgore1993}, in which we added an overall constant. The solid line is a fit to Eq. (\ref{steady_sliding}). {\bf (d)} $f(v)$ for paper \citep{Heslot1994}.}
 \label{fig:steadyState}
\end{figure}

Here we extend the recent ideas of \citet{Brener2002, Braun2009, Bouchbinder2011} into a realistic rate-and-state model of spatially extended frictional interfaces. As is well known, such interfaces are composed of an ensemble of contact asperities whose total area $A_r$ is much smaller than the nominal contact area $A_n$ and which exerts a shear stress $\tau$ that resists sliding motion. We decompose $\tau$ into an elastic part, emerging from the elastic deformations of contact asperities that are characterized by a coarse-grained stress $\tau^{el}$, and a viscous part $\tau^{vis}$
\begin{equation}
\label{eq:tauvisc}
\tau = \tau^{el}+ \tau^{vis}=  \tau^{el} + \eta \,v^* \! A  \,\, \mbox{sgn}(v) \log\left(1+\frac{|v|}{v^*}\right) \ ,
\end{equation}
where $\eta$ is a viscous-friction coefficient, $v$ is the slip velocity (slip rate), $v^*$ is a low-velocity cutoff scale and $A\!=\!A_r/A_n\!\ll\!1$ is the normalized real contact area. The viscous-stress $\tau^{vis}$, which increases with $v$ and scales with $A$, is usually associated with activated rate processes at asperity contacts (see also discussion in \citet{Bizzarri2011}). The $1$ inside the log ensures a regular behavior in the limit $v\to 0$, but otherwise plays no crucial role in what follows.

The next step is writing down a dynamic evolution equation for $\tau^{el}$. $\tau^{el}$ is stored at contact asperities at a rate determined by $v$ and that is proportional to both the interfacial elastic modulus $\mu_0$ and $A$. It is released as contact asperities are destroyed after slipping over a characteristic distance $D$ (of the order of the size of a contact as in conventional rate-and-state models \citep{Dieterich1979, Ruina1983, Baumberger2006}), when the asperity-level stress surpasses a yield-like threshold $\tau_c$. This physical picture is mathematically captured by writing \citep{Bouchbinder2011}
\begin{equation}
\label{eq:taudot}
\dot\tau^{el}  =  \mu_0 A \frac{v}{h} -\frac{\tau^{el} |v|}{D}\theta\left(\frac{\tau}{A}-\tau_c\right)\ ,
\end{equation}
where $h$ is the effective height of the asperities. Note that the coarse-grained stress $\tau$ is enhanced by a factor $A^{-1}\!\gg\!1$ at the asperities level and that the geometric nature of elastic stress relaxation, emerging from the multi-contact nature of the interface, is captured by the introduction of a spatial length $D$. The appearance of a Heaviside step function $\theta(\cdot)$ is an outcome of the basic notion of a local static threshold for sliding motion. The evolution law in Eq. (\ref{eq:taudot}) features a reversible elastic response at small shear stresses, $\tau \!\simeq\! \tau^{el} \!=\! \mu_0 A\,u/h$, where $u$ is the shear displacement. This elastic response is usually not included in friction models (but see \citet{Bureau2000, Shi2010}), even though it was directly measured experimentally \citep{Berthoud1998}.

To proceed, we write the normalized contact area $A$ in terms of a state variable $\phi$ as $A(\phi,\sigma)\!=\!A_0(\sigma)\left[1+b\log(1\!+\!\phi/\phi^*)\right]$ \citep{Baumberger2006}.
Here $\sigma$ is the (compressive) normal stress and $A_0(\sigma)\!=\!\sigma/\sigma_{\!H}$, where $\sigma_{\!H}$ is the hardness. The evolution of $A$ is phenomenologically captured by Dieterich's law \citep{Dieterich1979}, extended here by stipulating that the transition from the aging regime ($v\!=\!0$) to the sliding regime ($v\!\ne\!0$) is controlled by the same step function as in Eq. (\ref{eq:taudot}), yielding
\begin{equation}
 \label{eq:psidot}
  \dot\phi = 1 - \frac{\phi|v|}{D}\theta\left(\frac{\tau}{A}-\tau_c\right) \ ,
\end{equation}
where $\phi$ is interpreted as the ``geometric age'' of the contacts. Equations (\ref{eq:tauvisc})-(\ref{eq:psidot}) determine the evolution of $\tau(t)$, i.e. constitute our proposed friction law. We note that if Eq. (\ref{eq:taudot}) is replaced by its steady state solution, $\tau^{el}\!\sim\!A(\phi)$, our friction model becomes essentially identical to the conventional rate-and-state model (see also auxiliary material discussion).

Before we proceed we note a very important feature of rate-and-state friction models, which is not specific to the present model. In the absence of persistent sliding, $v\!=\!0$, we have $\phi\!=\!t$ and the contact area ages logarithmically $A\!\propto\!1+b\log(1\!+\!t/\phi^*)$, as is widely observed \citep{Baumberger2006}. The latter form suggests that the logarithmic law is cutoff at short timescales, smaller than $\phi^*$, as was directly confirmed experimentally in \citet{Dieterich1979, Nakatani2006, Ben-David2010b}. This very same short timescales cutoff manifests itself also under persistent sliding, $v\!\ne\!0$, for which we have $\phi\!=\!D/|v|$ and $A\!\propto\!1+b\log[1+D/(\phi^* |v|)]$. In this case, $A$ saturates at a finite value above a typical slip rate of order $D/\phi^*$ and the fixed-point of Eq. (\ref{eq:taudot}), $\tau^{el}\!\propto\!A$, becomes $v$-independent as well. As a consequence, $\tau$, which usually exhibits a velocity-weakening behavior at small $v$, becomes velocity-strengthening as the viscous-friction term in Eq. (\ref{eq:tauvisc}) takes over (see also discussion in \citet{Bizzarri2011}). Thus, rate-and-state friction models quantitatively {\bf predict} a non-monotonic dependence of the steady state sliding friction on the slip rate, an observation that has been largely overlooked in the literature (but see \citet{Weeks, Shibazaki2003, Baumberger2006, Yang2008}) and that will play an important role below.

\section{Steady State Rupture Fronts}

\begin{figure}[t]
 \centering
 \noindent\includegraphics[width=1\columnwidth]{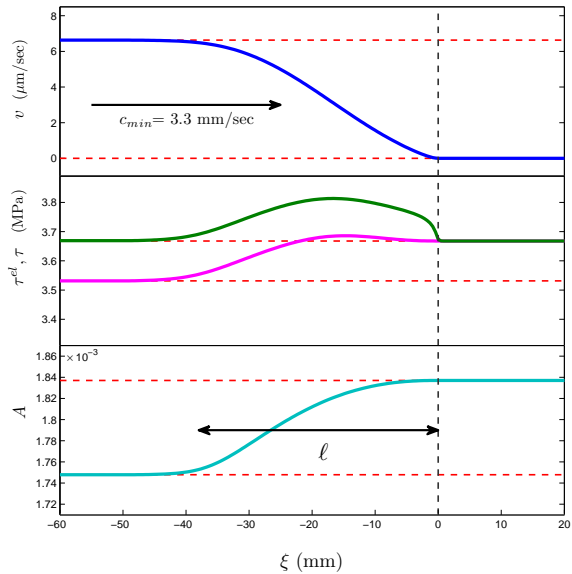}
 \caption{From top to bottom, $v(\xi)$, $\tau(\xi)$ (green), $\tau^{el}(\xi)$ (magenta) and $A(\xi)$ for a steady state rupture mode in granite propagating from left to right at $c_{min} \approx 3.3$ mm/sec $\ll c_s$. $\sigma\Delta f$ is the dynamic stress drop.}
 \label{fig:fronts}
\end{figure}

Propagating front solutions exist in multi-stable systems in which one homogeneous (space independent) solution invades another one, giving rise to non-trivial spatiotemporal structures. The spatially homogeneous solutions of Eqs. (\ref{eq:tauvisc})-(\ref{eq:psidot}), as a function of a driving stress $\tau^d$, are shown in Fig. \ref{fig:steadyState}a. A branch of elastic (static) solutions exists at $v\!=\!0$, where aging effects are neglected, i.e. we assume that $\psi_0\!\equiv\!b\log(1+t/\phi^*)$ is roughly constant for the timescales relevant for front propagation (essentially we set $t\!=\!\phi_0$). A branch of steady sliding solutions with $v\!>\!0$ takes the form
\begin{eqnarray}
\label{steady_sliding}
f= \frac{\tau_{ss}}{\sigma} \simeq f_0 +\alpha\log\left(1+\frac{v}{v^*}\right)+\beta\log\left(1+\frac{D}{\phi^* v}\right),
\end{eqnarray}
where $\alpha\!\equiv\!\eta\,v^*/\sigma_{\!H}$, $\beta\!\equiv\!\mu_0 D \,b/\sigma_{\!H} h$, $f_0\!\equiv\!\beta/b$
and $f$ is the steady sliding friction coefficient. Note that we neglected a term of order $\log^2$ in Eq. (\ref{steady_sliding}). As discussed above, steady sliding friction is indeed non-monotonic (when $\alpha \!<\! \beta$); friction is velocity-weakening for $v^* \!\ll\! v\! \ll\! D/\phi^*$ and velocity-strengthening for $v \!\gg\! D/\phi^*$, with a minimum at $v_m\!\simeq\!(D/\phi^*)(\beta-\alpha)/\alpha$.

At the minimum, we define the friction stress as $\tau_m=\tau_{ss}(v_m)$. Figs. \ref{fig:steadyState}b-d present experimental data for a polymer (PMMA), a rock (granite) and paper, where the last two data sets clearly demonstrate the non-monotonic nature of sliding friction, and the first one presumably does not span a sufficiently large range of $v$'s to detect a minimum.

Consider now a homogeneous driving stress $\tau^d$. For $\tau^d\!<\!\tau_m$ there exists only one stable homogeneous solution, the elastic (static) one. Upon increasing $\tau^d$ above $\tau_m$, three solutions exist: the elastic one with $v\!=\!0$ and two steady sliding solutions, one with $v\!<\!v_m$ (typically unstable) and one with $v\!>\!v_m$ (typically stable). The critical point $\tau^d\!=\!\tau_m$ corresponds to a bifurcation, which suggests a qualitative change in the behavior of the system. At this point we expect steady state propagating rupture, in which a solution with $v\!\ge\!v_m$ invades an elastic (static) solution with $v\!=\!0$, to emerge. Denote the propagation velocity of such fronts by $c$ and the one corresponding to $\tau^d\!=\!\tau_m$ by $c_{min}$.

In order to find propagating rupture solutions, and in particular to calculate $c_{min}$, we need to couple the friction law in Eqs. (\ref{eq:tauvisc})-(\ref{eq:psidot}) to an elastic body. It would not be easy to analytically calculate $c_{min}$ when the body is a 2D medium. Therefore, to gain analytic insight into the properties of the steady state fronts, we assume that the height $H$ of the elastic body (say in the $y$-direction) is much smaller than the spatial scale of variation $\ell$ of fields along the interface (in the $x$-direction), i.e. we consider a quasi-1D limit. Under these conditions we obtain (auxiliary material)
\begin{equation}
 \label{eq:udot}
  H\rho \,\partial_{tt} u(x,t)  \simeq  H\mu \,\partial_{xx} u(x,t) + \tau^d - \tau(x,t) \ ,
\end{equation}
where $\mu$ is the bulk shear modulus and $u$ is the interfacial shear displacement (slip) that satisfies $\partial_t u\!=\!v$. Note that we have omitted constants of order unity in Eq. (\ref{eq:udot}) and that in the quasi-1D limit both the driving stress $\tau^d$ and the friction stress $\tau$ do not appear as boundary conditions, but rather as terms in the ``bulk'' equation.

\begin{figure}
 \centering
 \noindent\includegraphics[width=1\columnwidth]{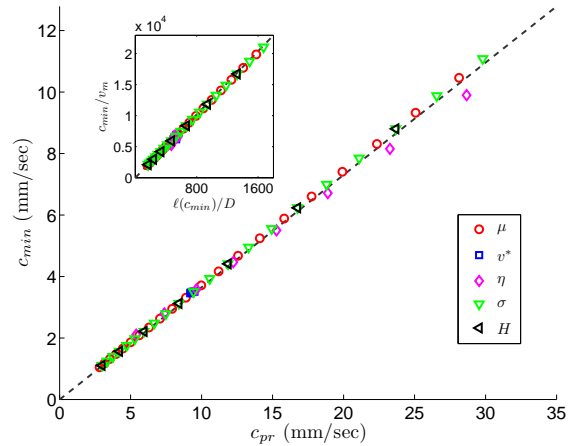}
 \caption{The numerically calculated $c_{min}$ for granite vs. the analytic prediction appearing on the right-hand-side of Eq. (\ref{eq:cScaling}), which we denote here as $c_{pr}$. (inset) $c_{min}/v_m$ vs. $\ell(c_{min})/D$ as obtained in the numerical calculations, cf. Eq. (\ref{eq:cScaling}). The dashed lines are guides to the eye.}
 \label{fig:scaling}
\end{figure}

We now look for steady state propagating solutions of Eqs. (\ref{eq:taudot}), (\ref{eq:psidot}) and (\ref{eq:udot}) in which all of the fields take the form $g(\xi\!=\!x-c\,t)$, where $c$ is the propagation velocity, such that a sliding solution at $\xi\!\to\!-\infty$ propagates into an elastic solution at $\xi\!\to\!\infty$. Smoothly connecting these two different solutions around $\xi\!=\!0$ provides solvability conditions that allow the calculation of $c$. We stress that $c$ must be distinguished from the slip rate $v$.

$c_{min}$ is being estimated using a scaling calculation in which the loading $\tau^d$ is homogeneous and equals to its threshold value $\tau_m$. A self-consistency constraint on the quasi-1D formulation is $H\!\ll\!\ell$, where $\ell$ is the spatial scale characterizing all of the fields in the front solution (as defined above). We first use $\partial_t\!=\!-c\partial_\xi$ to transform Eqs. (\ref{eq:taudot}), (\ref{eq:psidot}) and (\ref{eq:udot}) into the following set of coupled ordinary differential equations
\begin{eqnarray}
&&H\left(\mu/c - c \rho\right) \partial_\xi v = \tau^d-\tau \ , \label{eq:vdotSS}\\
&&-c \partial_\xi \tau^{el} = \mu_0\frac{v}{h} A(\phi,\sigma)-\frac{\tau^{el}|v|}{D}\theta\left(\frac{\tau}{A}-\tau_c\right),\label{eq:taudotSS}\\
&&-c\partial_\xi \phi  =  1-\frac{\phi|v|}{D}\theta\left(\frac{\tau}{A}-\tau_c\right) \ .\label{eq:psidotSS}
\end{eqnarray}
We stress that the front velocity $c$ in these equations is not a-priori known, but is rather a ``nonlinear eigenvalue'' of this problem, which is determined from the condition that the spatially-varying propagating solution properly converges to the homogeneous sliding solution at $\xi\!\to\!-\infty$ and to the homogeneous elastic solution at $\xi\!\to\!\infty$.

A scaling analysis of the above equations (auxiliary material) yields
\begin{eqnarray}
\label{eq:cScaling}
\ell(c_{min})\!\sim\!D\frac{c_{min}}{v_m} \quad\hbox{and}\quad c_{min} \sim \,v_m \sqrt{\frac{\mu\,H}{\sigma\,D\,\Delta f}} \ ,
\end{eqnarray}
where $\sigma\Delta f$ is the dynamic stress drop, cf. Fig. \ref{fig:fronts} (middle panel).

Several features of this central result are noteworthy. First, $c_{min}$ is finite and proportional to $v_m$. Second, it is independent of inertia, i.e. it does not scale with the elastic wave speed $c_s\!=\!\sqrt{\mu/\rho}\,$ \citep{Brener2002}. Finally, $c_{min}$ depends on: (i) the properties of the friction law, e.g. on constitutive parameters such as the viscous-friction coefficient $\eta$ (through $v_m$) and the (dimensionless) dynamic stress drop $\Delta f$, and on the microscopic geometric quantity $D$, (ii) the bulk geometry through $H$, (iii) the normal stress as $\sigma^{-1/2}$ and (iv) the bulk shear modulus $\mu$. We expect these features to remain qualitatively valid independently of the explicit form of the friction law and of dimensionality as long as steady sliding friction exhibits a non-monotonic behavior (cf. Fig. \ref{fig:steadyState}a), as suggested in \citet{Bouchbinder2011}.

To test the analytic prediction in Eq. (\ref{eq:cScaling}), we determine the friction parameters for a rock (granite) and a polymer (PMMA) using various sources and data sets (auxiliary material). In addition, we set $H\!=\!100\mu$m, and $\sigma\!=\!5$ MPa for granite (as in Fig. \ref{fig:steadyState}c) and $\sigma\!=\!1$ MPa for PMMA (as in \citet{Ben-David2010a, Ben-David2010b}). Finally, the state of the interface in the non-flowing region was chosen such that $\psi_0\!=\!0.06$ (granite) and $\psi_0\!=\!0.6$ (PMMA). In Fig. \ref{fig:fronts} we show a steady state rupture solution obtained by numerically integrating our model equations for granite. The propagation velocity, $c_{min}\!=\!3.3$mm/sec, is about than six orders of magnitude smaller than $c_s\!\sim\!10^3$m/sec, qualifying it as ``slow rupture'', and $\ell(c_{min})$ is on a mm scale, satisfying $H\!\ll\!\ell$ as required by self-consistency.

A similar calculation for PMMA (auxiliary material) yields $c_{min}\!=\!3.8$m/sec, which is about three orders of magnitude larger than $c_{min}$ for granite. This is expected since the square root term in Eq. (\ref{eq:cScaling}) is not dramatically different for the two materials, but $v_m$ is (cf. Figs. \ref{fig:steadyState}b-c). Recall that $v_m\!\sim\!D/\phi^*$ and that $D$ is in the $\mu$m scale for both materials (auxiliary material), which imply that the difference emerges from $\phi^*$. Indeed, $\phi^*\!\sim\!0.1\!-\!1$ sec for granite \citep{Dieterich1979, Nakatani2006} and $\phi^*\!\sim\!10^{-4}-10^{-3}$ sec for PMMA \citep{Ben-David2010b}.

We test the prediction in Eq. (\ref{eq:cScaling}) by each time varying one parameter on the right-hand-side and comparing the prediction to the numerically calculated $c_{min}$. The results are presented in Fig. \ref{fig:scaling} and exhibit excellent agreement between the analytic prediction and the numerically calculated values of $c_{min}$ for granite (similar results were obtained for PMMA). This result clearly and directly demonstrates the existence of friction-controlled slow rupture in our model.

\section{The Spectrum of Rupture Fronts}
\begin{figure}
 \centering
 \noindent\includegraphics[width=1\columnwidth]{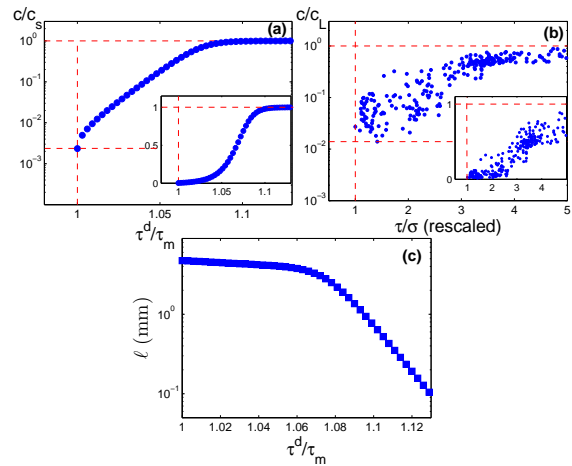}
 \caption{{\bf (a)} $c/c_s$ vs. $\tau^d/\tau_m$ for PMMA, under a fixed $\sigma$, in semi-log (main) and linear (inset) scales. {\bf (b)} $c/c_L$ vs. $\tau/\sigma$ in the PMMA experiments of \cite{Ben-David2010a} (courtesy of O. Ben-David and J. Fineberg) in semi-log (main) and linear (inset) scales. $c_L$ is the longitudinal wave-speed and $\tau/\sigma$ is rescaled such that the minimal value below which no rupture modes were observed equals unity. $c_{min}$ here is of the order of $10$ m/sec. {\bf (c)} $\ell$ vs. $\tau^d/\tau_m$ for the spectrum in panel {\bf (a)}.}
 \label{fig:fullSpectrum}
\end{figure}

The finite velocity scale $c_{min}$ implies there are no solutions with $c\!<\!c_{min}$, i.e. the existence of a ``forbidden'' range of velocities in the spectrum of steady state rupture modes \citep{Bouchbinder2011}. In Fig. \ref{fig:fullSpectrum}a we show the full spectrum of rupture propagation velocities as a function of $\tau^d\!\ge\!\tau_m$ for PMMA (a similar spectrum is obtained for granite, although $c_{min}$ is much smaller in this case). Indeed, there are no solutions with $c\!<\!c_{min}$ and there exists a continuum of states between $c_{min}$ and the elastic wave speed $c_s$. This continuous spectrum seems to be qualitatively similar to recent laboratory measurements \citep{Ben-David2010a}, reproduced here in Fig. \ref{fig:fullSpectrum}b. These measurements, though not obtained under globally homogeneous loading and were done in 2D, directly demonstrate the existence of a threshold driving stress, a minimal slow rupture velocity and saturation at an elastic wave speed. A detailed quantitative comparison to the experiments requires fully 2D calculations which are currently underway.

Upon increasing $\tau^d$ sufficiently above $\tau_m$, rupture travels at a non-negligible fraction of the sound speed and we can no longer neglect the inertial term in Eq. (\ref{eq:vdotSS}). A scaling analysis (auxiliary material) yields
\begin{equation}
\label{sharpening}
\ell(c\!\sim\!c_s)\sim e^{-\tau^d/ \alpha\,\sigma} \ll \ell(c_{min}) \ .
\end{equation}
The strong inequality results from the exponential decay of $\ell(c)$ with $\tau^d$ in the inertial regime and the typically small value of $\alpha$ ($\sim 0.01$). This result predicts that slow rupture is much less spatially localized as compared to ordinary fast rupture. In Fig. \ref{fig:fullSpectrum}c we test this prediction by plotting $\ell$ vs. $\tau^d/\tau_m$. The numerical results clearly confirm the theoretical prediction, demonstrating that indeed slow rupture is significantly less localized than rupture propagating at elastodynamic velocities. Furthermore, the exponential dependence predicted in Eq. (\ref{sharpening}) is quantitatively verified and the slope agrees with $-1/\alpha\,\sigma$.

\section{Summary and Conclusions}

Our results, based on a rate-and-state friction law, show that slow rupture is a well-defined and generic state of frictional interfaces. The non-monotonic dependence of the steady state sliding friction on the slip velocity gives rise to a new, friction-controlled, velocity scale $c_{min}$ below which no steady state rupture can propagate. Furthermore, our analysis demonstrates that rupture states span a continuum, from friction-controlled slow rupture to inertia-limited, earthquake-like, fast rupture \citep{Peng2010}. One may speculate that transient rupture modes observed under complex, spatially inhomogeneous, conditions are short-lived excitations of these steady rupture states, as was suggested within a specific context in \citet{Bouchbinder2011}. If true, steady state rupture fronts may play a role analogous to ``normal modes'' or ``eigenstates'' in other dynamical contexts.

The results presented are qualitatively consistent with recent laboratory measurements on PMMA \citep{Rubinstein2004, Ben-David2010a}, while similar results were obtained for a rock (granite). A quantitative comparison to experimental data requires 2D calculations which are currently underway. We hope to apply our ideas to a concrete geophysical system (e.g. to a slow/silent earthquake) in a future investigation.

\begin{acknowledgments}
We thank O. Ben-David and J. Fineberg for numerous insightful discussions. EB acknowledges support of the James
S. McDonnell Foundation, the Minerva Foundation with funding from the Federal German Ministry for Education and Research, the Harold Perlman Family Foundation and the William Z. and Eda Bess Novick Young Scientist Fund.
\end{acknowledgments}


\begin{thebibliography}{20}
\providecommand{\natexlab}[1]{#1}


\bibitem[{\textit{Baumberger and Berthoud}(1999)}]{Baumberger1999}
Baumberger, T., and P.~Berthoud  (1999), {Physical analysis of the state-and rate-dependent friction law. II. Dynamic friction}, \textit{Phys. Rev. B}, \textit{60}(6), 3928--3939.

\bibitem[{\textit{Baumberger and Caroli}(2006)}]{Baumberger2006}
Baumberger, T., and C.~Caroli  (2006), {Solid friction from stick--slip down to pinning and aging}, \textit{Advances in Physics}, \textit{55}(3-4), 279-348.

\bibitem[{\textit{Ben-David et~al.}(2010{\natexlab{a}})\textit{Ben-David,
  Cohen, and Fineberg}}]{Ben-David2010a}
Ben-David, O., G.~Cohen, and J.~Fineberg  (2010a), {The dynamics of the onset of frictional slip}, \textit{Science}, \textit{330}(6001), 211-214.

\bibitem[{\textit{Ben-David et~al.}(2010{\natexlab{b}})\textit{Ben-David, Rubinstein, and Fineberg}}]{Ben-David2010b}
Ben-David, O., S.~M. Rubinstein, and J.~Fineberg  (2010b), {Slip-stick and the evolution of frictional strength}, \textit{Nature}, \textit{463}(7277),  76-9.

\bibitem[{\textit{Berthoud and Baumberger}(1998)}]{Berthoud1998}
Berthoud, P., and T.~Baumberger  (1998), {Shear stiffness of a solid--solid multicontact interface}, \textit{Proc. R. Soc. London, Ser. A},
  \textit{454}(1974),  1615.

\bibitem[{\textit{Bizzarri}(2011)}]{Bizzarri2011}
Bizzarri, A. (2011), On the deterministic description of earthquakes, \textit{Rev. Geophys.}, 49, RG3002.

\bibitem[{\textit{Bouchbinder et~al.}(2011)\textit{Bouchbinder, Brener, Barel,
  and Urbakh}}]{Bouchbinder2011}
Bouchbinder, E., E.~A. Brener, I.~Barel, and M.~Urbakh  (2011), {Slow cracklike dynamics at the onset of frictional sliding}, \textit{Phys. Rev. Lett.}, \textit{107}(23), 235501.

\bibitem[{\textit{Braun et~al.} (2009)}]{Braun2009}
Braun, O., Barel, I., and Urbakh, M.  (2009), {Dynamics of transition from static to kinetic friction}, \textit{Phys. Rev. Lett.},  \textit{103}(1), 194301.

\bibitem[{\textit{Brener and Marchenko} (2002)}]{Brener2002}
Brener, E. A., and Marchenko, V. I.  (2002), {Frictional shear cracks}, \textit{J. Exp. Theor. Phys. Lett.}, \textit{76}, 211--214.

\bibitem[{\textit{Bureau et~al.} (2000)}]{Bureau2000}
Bureau, L., Baumberger, T., and Caroli, C.  (2000), {Shear response of a frictional interface to a normal load modulation},  \textit{Phys. Rev. E}, \textit{62}, 6810--6820.

\bibitem[{\textit{Dieterich}(1979)}]{Dieterich1979}
Dieterich, J.~H.  (1979), {Modeling of rock friction 1. Experimental results and constitutive equations},
  \textit{J. Geophys. Res.}, \textit{84}, 2161--2168.

\bibitem[{\textit{Heslot et~al.}(1994)\textit{Heslot, Baumberger, Perrin,
  Caroli, and Caroli}}]{Heslot1994}
Heslot, F., T.~Baumberger, B.~Perrin, B.~Caroli, and C.~Caroli  (1994), {Creep,  stick-slip, and dry-friction dynamics: Experiments and a heuristic model},  \textit{Phys. Rev. E}, \textit{49}(6), 4973--4988.

\bibitem[{\textit{Kato}(2003)}]{Kato2003}
Kato, N.   (2003), {A possible model for large preseismic slip on a deeper extension of
  a seismic rupture plane}, \textit{Earth Planet. Sci. Lett.},  \textit{216}(1-2), 17-25.

\bibitem[{\textit{Kilgore et~al.}(1993)}]{Kilgore1993}
Kilgore, B. D., Blanpied, M. L., and Dieterich, J. H.  (1993), {Velocity dependent friction of granite over a wide range of conditions}, \textit{Geophys. Res. Lett.}, \textit{20}(10), 903--906.

\bibitem[{\textit{Liu and Rice}(2005)}]{Liu2005a}
Liu, Y., and J.~Rice  (2005), {Aseismic slip transients emerge spontaneously in  three-dimensional rate and state modeling of subduction earthquake sequences}, \textit{J. Geophys. Res.}, \textit{110}(B8),  1-14.

\bibitem[{\textit{Nakatani and Scholz}(2006)}]{Nakatani2006} Nakatani, M., and Scholz, C. H.   (2006), {Intrinsic and apparent short-time limits for fault healing: theory, observations, and implications for velocity-dependent friction}, \textit{J. Geophys. Res.} \textit{111}, B12208.

\bibitem[{\textit{Nielsen et~al.}(2010)\textit{Nielsen, Taddeucci, and  Vinciguerra}}]{Nielsen2010}
Nielsen, S., J.~Taddeucci, and S.~Vinciguerra (2010), {Experimental observation of stick-slip instability fronts}, \textit{Geophys. J. Int.},  \textit{180}(2), 697--702.

\bibitem[{\textit{Peng and Gomberg}(2010)}]{Peng2010}
 Peng, Z., and Gomberg, J.  (2010), {An integrated perspective of the continuum between earthquakes and slow-slip phenomena}, \textit{Nat. Geosci.}, \textit{3}(9), 599--607.

\bibitem[{\textit{Rubinstein et~al.}(2004)\textit{Rubinstein, Cohen, and  Fineberg}}]{Rubinstein2004}
Rubinstein, S.~M., G.~Cohen, and J.~Fineberg   (2004), {Detachment fronts and the onset  of dynamic friction}, \textit{Nature}, \textit{430}(August), 1005-1009.

\bibitem[{\textit{Ruina}(1983)}]{Ruina1983}
Ruina, A.   (1983), {Slip instability and state variable friction laws}, \textit{J. Geophys. Res.}, \textit{88},10359--10370.

\bibitem[{\textit{Shi et~al.}(2010)}]{Shi2010}
Shi, Z., Needleman, A., and Ben-Zion, Y.  (2010), {Slip modes and partitioning of energy during dynamic frictional sliding between identical elastic-viscoplastic solids}, \textit{Int. J. Fract.}, \textit{162},51-67.

\bibitem[{\textit{Shibazaki and Iio}(2003)}]{Shibazaki2003}
Shibazaki, B., and Y. Iio  (2003), {On the physical mechanism of silent slip events along the deeper part of the seismogenic zone}, \textit{Geophys. Res. Lett.}, \textit{30}(9), 1489.

\bibitem[{\textit{Weeks}(1993)}]{Weeks}
Weeks, J. D.  (1993), {Constitutive laws for high--velocity frictional sliding
and their influence on stress drop during unstable slip}, \textit{J. Geophys. Res.},
\textit{98}, 17637--17648.

\bibitem[{\textit{Yang et~al.}(2008)}]{Yang2008}
Yang, Z., Zhang, H. P., and Marder, M.  (2008), {Constitutive laws for high-velocity frictional sliding and their influence on stress drop during unstable slip}, \textit{Proc. Natl. Acad. Sci. U.S.A.}, \textit{195}, 13264--13268.

\bibitem[{\textit{Yoshida and Kato}(2003)}]{Yoshida2003}
Yoshida, S., and N. Kato  (2003), {Episodic aseismic slip in a two--degree--of--freedom block--spring model}, \textit{Geophys. Res. Lett.},
\textit{30}(13),  1681.

\end{thebibliography}

\begin{thebibliography}{99}
\providecommand{\natexlab}[1]{#1}

\bibitem{Baumberger2006A} Baumberger, T. \& Caroli C. {Solid friction from stick-slip down to
  pinning and aging}, \textit{Adv. Phys.}, \textbf{55}, 279--348 (2006).

  \bibitem{Bouchbinder2011A}
Bouchbinder, E., Brener E. A., Barel I. \& Urbakh M., Slow cracklike dynamics at the onset of frictional sliding, \textit{Phys. Rev. Lett.}, \textbf{107}, 235501 (2011).


\bibitem{Berthoud1998A} Berthoud, P. \& Baumberger T., Shear stiffness of a solid-solid
  multi-contact interface, \textit{Proc. Roy. Soc. Lon. A},
  \textbf{454}(1974), 1615-1634, (1998).


\bibitem{Rubinstein2004A} Rubinstein, S. M., Cohen G. \& Fineberg J. Detachment fronts and the onset
  of dynamic friction, \textit{Nature}, \textbf{430}, 1005-1009, 2004.

\bibitem{Ben-David2010aA}
Ben-David, O., Cohen G. \& Fineberg J. The dynamics of the onset of
  frictional slip, \textit{Science}, \textbf{330}, 211-214, (2010).

\bibitem{Scholz1989A}
Yoshioka, N. \& Scholz, C., Elastic properties of contacting surfaces under normal and shear loads 2. comparison of theory with experiment
  , \textit{J. Geophys. Res.}, \textbf{94}, 17,691-17,700, (1989).

\bibitem{SanoA}
Sano, O., Kudo, Y., \& Mizuta Y., Experimental determination of elastic constants of Oshima granite, barre granite, and Chelmsford Granite,
\textit{J. Geophys. Res.}, \textbf{97}, 3367-3379, (1992).

\bibitem{Dieterich1978A}
Dieterich J., Time-dependent friction and the mechanics of stick-slip, \textit{Pure and Applied Geophys.}, \textbf{116}, 790-806, (1978).

\bibitem{Rice2001A}
Rice, J., Lapusta, N., \& Ranjith, K, Rate and state dependent friction and the stability of sliding between elastically deformable solids, \textit{J. Mech. Phys. Solids}, \textbf{49}, 1865-1898, (2001).

\bibitem{Ben-David2010bA} Ben-David, O., Rubinstein S. M. \& Fineberg J. Slip-stick and the evolution
  of frictional strength. \textit{Nature}, \textbf{463}, 76-9, (2010).

\bibitem{Bureau2000A} Bureau, L., Baumberger T. \& Caroli C., Shear response of a frictional interface to a normal load modulation,
  \textit{Phys. Rev. E}, \textbf{62}, 6810--6820, (2000).

\end{thebibliography}
\bibliographystyle{agufull}

\onecolumngrid
\newpage
\renewcommand{\theequation}{A\arabic{equation}}
\renewcommand{\thefigure}{S\arabic{figure}}
\renewcommand{\thesection}{S\arabic{section}}
\setcounter{section}{0}
\setcounter{equation}{0}
\vspace{2cm}
\begin{center}
{\Large{\bf Supplementary Material}}\vspace{1cm}
\end{center}	

\twocolumngrid

\section{The proposed friction law}

The basic physical idea behind the proposed friction law is that contact asperities experience elastic deformation that contributes to
the friction stress $\tau$. We denote this coarse-grained elastic contribution by $\tau^{el}$. We then write $\tau$ as a sum of $\tau^{el}$ and
a viscous-friction stress $\tau^{vis}$, $\tau \!=\!\tau^{el}+\tau^{vis}$. The latter increases with $v$ and vanishes as $v\!\to\!0$. It is a material (constitutive) property which is not directly related to the multi-contact nature of the interface, though it must be proportional to the amount of contact. This picture is analogous to the standard Kelvin-Voigt model of visco-elasticity, in which an elastic spring and a viscous dashpot element are connected in parallel.
In order to deviate as little as possible from conventional modeling, we write
\begin{equation}
 \tau^{vis}(v,A(\sigma ,\phi))  =  \eta \,v^* A(\sigma ,\phi) \log\left(1+\frac{|v|}{v^*}\right) \ ,
\end{equation}
which is essentially the logarithmic ``direct effect'' term used in classical rate-and-state models \cite{Baumberger2006A}. $\eta$ is a viscous-friction coefficient of stress/velocity dimension and $v^*$ is the same slip rate scale as in (1). The $1$ inside the log ensures a regular behavior in the limit $v\!\to\!0$, but otherwise plays no
central role here.

The next step is writing down a dynamic evolution equation for $\tau^{el}$ \cite{Bouchbinder2011A}.
When the shear stress is small, we expect a purely elastic response. Therefore, the elastic strain rate experienced by a population of contact asperities of an effective height $h$ reads
\begin{equation}
\frac{v^{el}}{h} = \frac{\dot\tau^{el}}{A \mu_0} \ , \label{eq:v}
\end{equation}
where $v^{el}\!=\!\dot{u}^{el}$ represents here the time derivative of an elastic (reversible) shear displacement and $\mu_0$ is an interfacial shear modulus. Assuming that $A$ is time-independent (or a slowly varying function of time), we obtain
\begin{equation}
\tau^{el} = A \mu_0 u^{el}/h \ , \label{eq:u}
\end{equation}
which was directly observed experimentally \cite{Berthoud1998A}. These experiments allow us to constrain the ratio $\mu_0/h$.

When the stress at the level of contact asperities reaches a material dependent strength parameter $\tau_c$, irreversible slip initiates. After an inelastic slip over the typical linear size of contact asperities $D$, contact asperities are being destroyed (i.e. lose contact) and release their elastic stress. This physical picture in which inelastic slip is initiated upon surpassing a threshold, leading to a relaxation of the elastic stress during a timescale determined by $D/|v|$, is analogous to elasto-plastic behavior of bulk solids. Mathematically, it reads \cite{Bouchbinder2011A}
\begin{equation}
\label{eq:taudotA}
\dot \tau^{el}  =  \frac{\mu_0 A(\sigma, \phi)}{h} v -\frac{\tau^{el} |v|}{D}\theta\left(\frac{\tau}{A}-\tau_c\right) \ ,
\end{equation}
where the stress relaxation term (second term on the right-hand-side) operates only when $\tau\!>\!A\tau_c$. Note that the coarse-grained (macroscopic) stress $\tau$ is enhanced by a factor $A^{-1}\!\ll\!1$ at the asperities level,
accounting for the fact that a dilute population of contact asperities carries the macroscopic stress.
For $\tau\!<\!A\tau_c$ we recover the measurable elastic response with no persistent sliding.
For $\tau\!>\!A\tau_c$, we have $\tau^{el}\!=\!A\mu_0 D/h$ (in steady state) which depends on $v$ only through $A$. This happens because $v$ controls both the rate of elastic loading and of inelastic relaxation.

To complete the formulation of our friction law, we need an evolution equation for $A(\sigma, \phi)$, where $\sigma$ is the compressive normal stress and $\phi$ is a state variable. For that aim we write \cite{Baumberger2006A}
\begin{equation}
\label{eq:psiDefiniton}
 \begin{split}
   A(\sigma, \phi)&=A_0(\sigma)\left[1+\psi(\phi)\right],\\[2mm]
   \psi&\equiv b\log\left(1+\frac{\phi}{\phi^*}\right) \ ,
\end{split}
\end{equation}
where $A_0(\sigma)\!=\!\sigma/\sigma_{\!H}$ ($\sigma_{\!H}$ is the hardness) is an instantaneous equation of state. The evolution of $A$ is then determined by the evolution of $\phi$, which is given in Eq. (3) in the main text
\begin{equation}
 \label{phidotA}
  \dot\phi = 1 - \frac{\phi\,|v|}{D}\theta\left(\frac{\tau}{A}-\tau_c\right) \ .
\end{equation}

By comparing Eqs. (\ref{eq:taudotA}) and (\ref{phidotA}) we note a few points. First, since the evolution equation for $\tau^{el}$ depends on $A(\sigma,\phi)$, there exists some interdependence between $\tau^{el}$ and the state variable $\phi$. This implies that the relative time scales of variation of $\tau^{el}$ and $\phi$ may give rise to different dynamics. In Eqs. (\ref{eq:taudotA}) and (\ref{phidotA}) above we assumed that $\tau^{el}$ and $\phi$ evolve on the same timescale $D/v$. If, on the other hand, $\tau^{el}$ evolves much faster than $\phi$, we can ``integrate out'' the $\tau^{el}$ dynamics and essentially recover the classical rate-and-state model \cite{Baumberger2006A}. This is discussed in Sec. \ref{classical_RSF}.

Finally, we comment on another feature of the steady state sliding curve described by Eqs. (4) in the main text. This curve diverges logarithmically
as $v\!\to\!0$ since it contains a term proportional to $\log[1+D/(\phi_0|v|)]$. We expect this unphysical divergence to be regularized at very small
slip rates, when competing slow timescales become relevant. We do not
consider such effects in the present paper.

\section{The quasi-1D limit of the momentum balance equation}

Here we show how to derive Eq. (5) in the main text. The bulk force balance equation reads
\begin{equation}
\label{momentum}
\rho \ddot{\B u} = \nabla \cdot \B \sigma \ ,
\end{equation}
where $\rho$ is the mass density, $\B u$ is the displacement field and $\B \sigma$ is the stress tensor field. We assume the bulk is linear elastic, i.e.
\begin{equation}
\label{Hooke}
\B \sigma = \lambda \, \operatorname{tr}(\nabla \B u) + \mu \left[(\nabla \B u)+(\nabla \B u)^T \right] \ ,
\end{equation}
where $\lambda$ is the first Lam\'e coefficient and $\mu$ is the shear modulus.
Consider then a long linear elastic strip of height $H$ in frictional contact with a semi-infinite half plane at $y\!=\!0$ ($x$ is the coordinate parallel to the interface). For simplicity we ignore the third dimension $z$. The external boundary conditions at $y\!=\!H$ are chosen here to be the normal and tangential stresses, $\sigma_{yy}(x,y\!=\!H,t)$ and $\sigma_{xy}(x,y\!=\!H,t)$. The friction law, Eq. (1) in the main text, dictates the shear stress at the interface, $\sigma_{xy}(x,y\!=\!0,t)\!=\tau(x,t)$.

In order to obtain the quasi-1D limit of this 2D formulation we assume that $H$ is much smaller than any lengthscale $\ell$ that characterizes the variation of quantities in the $x$-direction. For simplicity we also choose $u_y(x,y\!=\!0,t)\!=\!0$. In the leading approximation with respect to $H/\ell$, the solution takes the form $u_x(x,t)$ and $u_y\!\sim\!y$, which automatically satisfies the $y$-component of Eq. (\ref{momentum}) (with Eq. (\ref{Hooke})). We then integrate the $x$-component of Eq. (\ref{momentum}) over $y$ from $0$ to $H$, obtaining
\begin{equation}
H \rho \,\ddot{u}_x \simeq H \,\pa_x \sigma_{xx} + \sigma_{xy}(x,H,t)- \sigma_{xy}(x,0,t) \ ,
\end{equation}
Finally, we define $\tau^d\!=\!\sigma_{xy}(x,H,t)$ and use Eq. (\ref{Hooke}) to obtain $\sigma_{xx}\!\sim\!\mu \pa_x u_x$ (recall that $\lambda\!\sim\!\mu$), which leads to
\begin{equation}
H \rho \,\ddot{u}_x \simeq H \mu \,\pa_{xx} u_x + \tau^d - \tau(x,t) \ ,
\end{equation}
which is Eq. (5) in the main text. Such a reduction from 2D to 1D was discussed in \cite{Bouchbinder2011A}. We finally stress that the quasi-1D approximation breaks down when $\ell \simeq H$, which does not necessarily imply unrealistically small values of $H$.

\section{Predictions for $c_{min}$ and $\ell(c)$}

Here we show how to derive the prediction for $c_{min}$ and $\ell(c)$ in Eqs. (9)-(10) in the main text using Eqs. (6)-(8), in the spirit of \cite{Bouchbinder2011A}. We focus on the minimum of the steady sliding curve,
$\tau^d\!=\!\tau_m$ and $v\!=\!v_m$, and denote the spatial lengthscale of variation of the rupture fields at this point by $\ell(c_{min})$. The typical timescale of variation of the fields is $D/v_m$. The passage time of the front is $\ell(c_{min})/c_{min}$. Equating the two yields
\begin{equation}
 \ell(c_{min}) \sim D \frac{c_{min}}{v_m} \ . \label{eq:clv}
\end{equation}
To proceed, we estimate the left-hand-side of Eq. (6) by $- H \mu v_m/[\ell(c_{min}) c_{min}]$, where we neglected inertia, assuming $c_{min}\!\ll\!c_s\!=\!\sqrt{\mu/\rho}$.
We estimate the right-hand-side, $\tau^d-\tau^{el}-\tau^{vis}$, as the dynamic stress drop $\Delta\tau\!=\!\sigma\Delta f$ (i.e. the difference between the peak stress and the steady sliding stress, cf. Fig. 2, middle panel). For the latter we use
\begin{eqnarray}
\hspace{-0.55cm}
 \tau^d=\tau_m &\approx& A_0 \left[\frac{\mu_0 D}{h} +\eta \,v^* \log \left(1+\frac{v_m}{v^*}\right)\right],\label{taum}\\
 \tau^{el} &\approx& \tau_c\,A_0(1+\psi_0) \ ,\label{tauel}\\
 \tau^{vis} &\approx& \eta\,v^* A_0 \log\left(1+\frac{v_m}{v^*}\right) \ , \label{tauvis}
\end{eqnarray}
which leads to
\begin{equation}
\Delta\tau=\sigma\Delta f \approx \frac{\sigma}{\sigma_{\!H}}\left[\tau_c (1+\psi_0) - \frac{\mu_0 \,D}{h}\right] \ .
\end{equation}
Substituting these estimates in Eq. (6) gives
\begin{equation}
-\frac{\mu H v_m}{\ell \,c_{min}} \approx -\sigma\Delta f \ ,
\end{equation}
which immediately yields Eq. (9) in the main text when Eq. (\ref{eq:clv}) is used.

A few comments regarding the choice of the scaling estimates in Eqs. (\ref{taum})--(\ref{tauvis}) are in place. These three terms estimate the amount by which $\tau^{el}+\tau^{vis}$ overshoots $\tau^d$ (the dynamic stress drop), which drives the $v$ variation on the left-hand-side. $\tau^d$ is estimated in Eq. (\ref{taum}), where we neglected the steady-state value of $\psi\!=\!b\log(1+\phi/\phi^*)$, as it is much smaller than unity. In order to obtain a very rough estimate of the peak value of $\tau^{el}+\tau^{vis}$, we approximated $\tau^{el}$ in Eq. (\ref{tauel}) by its value when the interface starts to break and $\tau^{vis}$ in Eq. (\ref{tauvis}) was estimated by its value in the sliding region (again neglecting the steady state value of $\psi$). This rough estimate for the peak value of $\tau^{el}+\tau^{vis}$ may not always be accurate, and significant changes of the constitutive parameters might lead to a somewhat different choice of scaling estimates. However, these estimates and some possible variants of them, properly capture the essence of the scaling properties of $c_{min}$.

Upon increasing $\tau^d$ sufficiently above $\tau_m$, rupture travels at a non-negligible fraction of the sound speed and we can no longer neglect the inertial term in Eq. (6).
The sliding velocity in the sliding region is the solution corresponding to larger $v$ of two solutions of the equation (see Fig. 1a in the main text)
\begin{equation}
\frac{\tau^d}{\sigma} = \frac{\beta}{b} + \alpha \log\left(1+\frac{v}{v^*}\right)+\beta \log\left(1+\frac{D}{\phi^* v}\right)\ .
\end{equation}
For $v\!\gg\!v_m$ we can neglect the second logarithmic term and solve for $v$
\begin{equation}
\label{v_inertial}
v\sim \exp\left[\frac{\tau^d}{\sigma\,\alpha} - \frac{\beta}{b\,\alpha}\right]v^* \ .
\end{equation}
The analog of relation (\ref{eq:clv}) for the inertial regime reads $\ell(c\!\sim\!c_s) \!\sim\! D c/v$. This, together with Eq. (\ref{v_inertial}), yields Eq. (10) in the main text.


\section{Material parameters for granite and PMMA}

Frictional parameters seem to be sensitive to environmental and experimental conditions.
Somewhat different results for (supposedly) the same material, sometimes by the same research group, were reported. However, the trends are robust, as well as the order of magnitude of the parameters.
With this in mind, we compiled a list of parameters with which our model satisfactorily describes various data sets for granite and PMMA. Granite is a rather representative crustal rock and
PMMA was extensively characterized in laboratory measurements, including the interfacial elastic response \citep{Berthoud1998A}. Moreover, PMMA was used in the most conclusive experiments which demonstrated
the existence of slow rupture modes \citep{Rubinstein2004A, Ben-David2010aA}.
\begin{table}[here]
 \begin{tabular}{|c|c||c|c|c}
  \hline
 $\mu$ & 20 GPa &  $\tau_c$ & 2 GPa\\ \hline
 $\sigma_{\!H}$ & 2.86 GPa &  $b$ & 0.0071\\ \hline
 $\rho$ & 2,600 Kg/m$^3$ & $D$ & 5 $\mu$m  \\ \hline
 $v^*$ & 0.0035 $\mu$m/sec & $D/\phi^*$ & 16 $\mu$m/sec  \\ \hline
 $\eta$ & 2.91 GPa ($\mu$m/s)$^{-1}$ & $\mu_0$/h& 400 MPa/$\mu$m\\ \hline
 \end{tabular}
  \caption{Elastic and frictional parameters for granite.}
 \label{tab1}
\end{table}

We extracted the material parameters of granite from various sources. $\mu\!=\!20$GPa and $\rho\!=\!2,600$kg/m$^3$ were taken in accordance with \cite{Scholz1989A,SanoA}. $D\!=\!5\mu$m was taken from Fig. 5 of \cite{Dieterich1978A}. $H$ was taken to be $100\mu$m, as reported in the main text, to ensure $H\!\gg\!D$. We then used Eq. (4) in the main text to fit the steady-state sliding friction coefficient (Fig. 1c in the main text). This fit yielded the values of $\alpha,\beta,D/\phi^*$ and $v^*$. Using these parameters and the definitions $\alpha\!=\!\eta v^*/\sigma_H,\beta\!=\!\mu_0D b/\sigma_H h$ we extracted $\phi^*,\sigma_H,\eta$, and $b$, all of which are summarized in Table \ref{tab1}. The value  $\phi^*\!=\!0.31$sec is consistent with \cite{Dieterich1978A}. $\tau_c$ was estimated as $0.1\mu=2$GPa, as suggested in \cite{Rice2001A}. Lacking experimental measurements of interfacial elasticity in granite, we estimated $\mu_0/h$ to be roughly $\tau_c/D$, which gives reasonable results. As a self-consistency check of the latter estimate, we note that our parameters imply $\mu_0 D/h\!=\!2$GPa. $D/h$ was estimated in \cite{Dieterich1978A} as $0.2-0.5$, yielding $\mu_0=4-10$GPa. This suggests that the interfacial elastic modulus is somewhat softer than the bulk one, though of the same order of magnitude.

The material parameters of PMMA were likewise extracted from a variety of sources. The interfacial elastic response data of Fig. 2 in \citep{Berthoud1998A} indicates that $h \sigma_{\!H}/\mu_0$ is in the $\mu$m scale.
The slope of the ageing data in Fig. 9 of \citep{Baumberger2006A} implies $\beta\!=\!\frac{\mu_0 D \,b}{\sigma_{\!H} h}\!\simeq\!0.02$.
The ``direct effect'' measurement in Fig. 4 of \citep{Baumberger2006A} implies $\alpha=\eta v^*/\sigma_{\!H}\!\simeq\!0.005$. $v^*$ was estimated as the lowest slip rate for which the logarithmic ``direct effect'' is observed.
The data presented in Fig. 1b in the main text is consistent with $\beta\!-\!\alpha\!\simeq\!0.015$. $D$ was determined to be $0.9\mu$m in \cite{Ben-David2010bA} and $0.4\mu$m in \cite{Bureau2000A}.
With these constraints, and using known values of independently measured parameters such as $\mu$, $\sigma_{\!H}$, $\rho$ and $\tau_c$
(which is estimated as the yield stress), we fitted all of the aforementioned experimental data. The parameters are summarized in Table \ref{tab2}.

\begin{table}[here]
 \begin{tabular}{|c|c||c|c|c}
  \hline
 $\mu$ & 3.1 GPa &  $\tau_c$ & 130 MPa\\ \hline
 $\sigma_{\!H}$ & 540 MPa &  $b$ & 0.075\\ \hline
 $\rho$ & 1,200 Kg/m$^3$ & $D$ & 0.5 $\mu$m  \\ \hline
 $v^*$ & 0.1 $\mu$m/sec & $D/\phi^*$ & 1.5 mm/sec  \\ \hline
 $\eta$ & 27 MPa ($\mu$m/s)$^{-1}$ & $\mu_0$/h& 300 MPa/$\mu$m\\ \hline
 \end{tabular}
  \caption{Elastic and frictional parameters for PMMA.}
 \label{tab2}
\end{table}

\begin{figure}
  \centering
  \noindent\includegraphics[width=1.02\columnwidth]{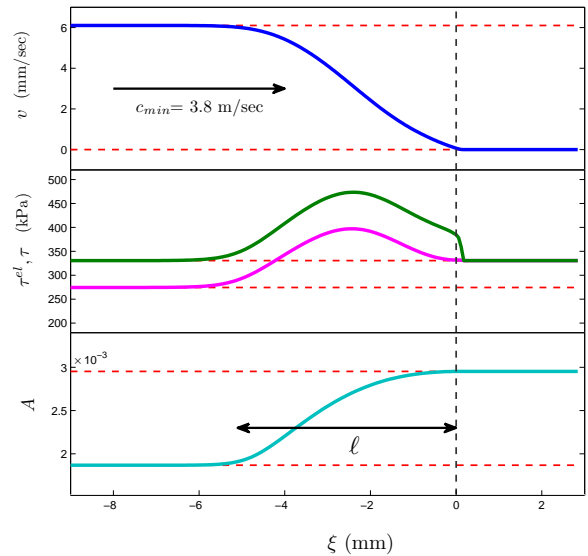}
  \caption{From top to bottom, $v(\xi)$, $\tau(\xi)$ (green), $\tau^{el}(\xi)$ (magenta) and $A(\xi)$ for a steady state rupture mode propagating in PMMA from left to right at $3.8$ m/sec $\ll c_s$.
  The spatial scale of variation of the fields is denoted by $\ell$. The homogeneous states that the rupture mode smoothly connects are marked by horizontal dashed lines.}
  \label{fig:frontsA}
 \end{figure}

In Fig. 2 in the main text we present the solution of Eqs. (6)-(8) for granite with $\tau^d\!=\!\tau_m$. For completeness, we present here in Fig. \ref{fig:frontsA} the corresponding solution for PMMA.
The major quantitative difference is the value of $c_{min}$, which is discussed in the main text.

\section{2D formulation}

In the main text a quasi-1D model was studied. Looking forward to future applications in 2D and 3D (currently underway) we briefly present here such formulations,
focusing on the boundary equations imposed by the friction law in both the elastic and sliding regimes.
We first discuss the elastic response below the threshold. For that aim, consider a block of height $H+h$ and elastic shear
modulus $\mu$, under a uniform shear stress $\tau$ at the top. Consider then an interfacial boundary layer of shear modulus $\mu_0$ and height $h$.
Denote the the displacement in the x-direction at $y\!=\!H\!+\!h$ by $u_x(H)$ and
the displacement at $y\!=\!h$ by $u^{el}$. We immediately obtain
\begin{equation}
\tau = \mu (u(H)-u^{el})/H,\quad \tau = \mu_0 A u^{el}/h \ .
\end{equation}
Solving for $u_x(H)$ and $u^{el}$ we obtain
\begin{equation}
u = \frac{H \tau}{\mu_0}\left(\frac{\mu_0}{\mu}+\frac{h}{H\,A} \right),\quad u^{el} =  \frac{h \,\tau}{\mu_0 A} \ .
\end{equation}
For
\begin{equation}
\label{ineq}
\frac{\mu_0}{\mu} \gg \frac{h}{H\,A} \ ,
\end{equation}
we have $u^{el} \!\ll\! u_x(H)$, motivating the boundary condition $u_x(x,y\!=\!0,t) \!\approx\! 0$, which more generally reads
$\dot{u}_x(x,y\!=\!0,t)\!\approx\!0$.
In the experiments of \cite{Berthoud1998A} the inequality (\ref{ineq}) was not satisfied, even though a macroscopic system was studied,
since $A\!\sim\!\sigma$ was relatively low. Then, interfacial elasticity was directly observed. In addition, in our quasi-1D calculations $\sigma$
was large, but $H$ was small, again invalidating the inequality. We believe, however, that the inequality in (\ref{ineq}) is satisfied for
macroscopic systems ($h$ in our case is of the order of $1\!-\!10 \mu$m) under large normal stresses ($\sim 1$MPa or more)
and rather generally leads to the boundary condition $\dot{u}_x(x,y\!=\!0,t) \!\approx\! 0$ below the threshold. Above the threshold ($\tau\!>\!\tau_c\,A$), in the sliding regime, Eqs. (2)-(3) in the main text are used with $\theta(\cdot)\!=\!1$. This is a stress-controlled boundary condition, as opposed to a displacement-controlled boundary condition below the threshold.

\section{Relation to conventional rate-and-state models}
\label{classical_RSF}

Finally, we note that the conventional rate-and-state model \cite{Baumberger2006A} features similar slow rupture solutions. To see this, replace Eq. (2) in the main text with its sliding steady state solution,
$\tau^{el}\!\sim\!A(\phi)$ (which means that $\tau^{el}$ evolves much faster than $\phi$) , and set $\theta(\cdot)\!=\!1$ in Eq. (3). The result is the classical formulation of rate-and-state friction \cite{Baumberger2006A}
\begin{equation}
\label{RSF}
 \begin{split}
 \hspace{-0.5cm} f(v,\phi) \!&=\! f_0 \!+\!\alpha\log\left(1+\frac{v}{v^*}\right)\!+\!\beta\log\left(1+\frac{\phi}{\phi^*}\right),\\
\hspace{-0.5cm} \dot\phi &= 1 - \phi\,|v|/D \ ,
\end{split}
\end{equation}
where $f(0,\phi)\sigma$ is understood as a threshold for the onset of slip. Then, the analog of Eqs. (6)-(8) in the main text read
\begin{eqnarray}
&&H\left(\mu/c - c \rho\right) \partial_\xi v(\xi) = \tau^d-f(v,\phi)\sigma\ , \label{eq:vdotSS_A}\\
&&-c\,\partial_\xi \phi(\xi)  =  1-\phi(\xi)\frac{|v|}{D} \ .\label{eq:phidotSS_A}
\end{eqnarray}
\begin{figure}
 \centering
 \noindent\includegraphics[width=1.02\columnwidth]{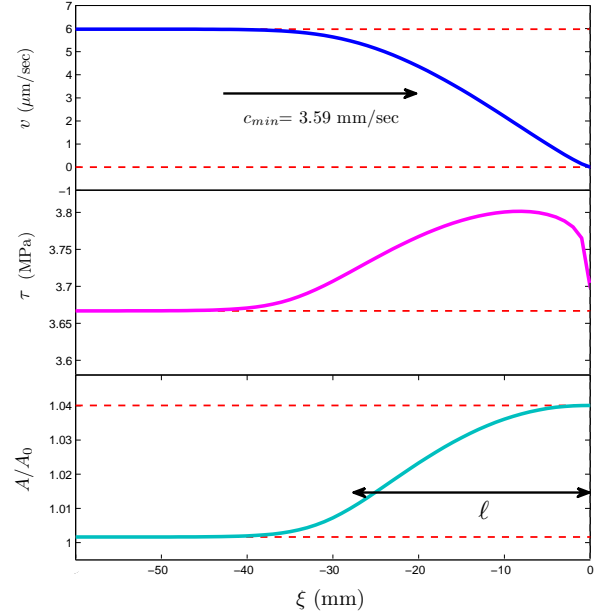}
 \caption{From top to bottom, $v(\xi)$, $\tau(\xi)$ and $A(\xi)/A_0=1+b\log\left[1+\phi(\xi)/\phi^*\right]$ for a steady state rupture mode propagating in granite from left to right at $3.59$ mm/sec, using the classical rate-and-state model in
 Eqs. (\ref{eq:vdotSS_A})-(\ref{eq:phidotSS_A}). $b=0.0071$ as in Table \ref{tab1} and $\xi>0$ is not shown since we did not solve anything in this range. Compare these results to Fig. 2 in the main text.}
 \label{fig:frontsB}
\end{figure}

These equations are valid for in the flowing region, $\xi\!<\!0$, while one should not solve for $\xi\!>\!0$,
but rather set the boundary conditions at $\xi\!=\!0$ to $v(0)\!=\!0$ and $\phi(0)\!=\!\phi_0$. Integrating these equations
with $\phi_0\simeq 10^3$ sec, $f_0=0.7$, $\alpha=3.6\! \times\! 10^{-3}$ and $\beta=5\! \times\! 10^{-3}$, and determining $c$
from the solvability condition of converging to the steady
sliding fixed point as $\xi \to -\infty$, we obtain the solution presented in Fig. 2.

Comparing this figure to Fig. 2 in the main text, we
see that the results agree semi-quantitatively with one
another. Therefore, we conclude that conventional rate-and-state friction models feature slow rupture solutions
as long as they exhibit a non-monotonic steady sliding behavior.
Furthermore, while the dynamics of $\tau^{el}$ in Eq. (2) in the main text are physically motivated and supported by
measurements of interfacial elastic response, this equation does not play a crucial mathematical role in obtaining slow rupture solutions.

\end{document}